\documentclass[ 
                breakaddress,  
                proof,         
]{nrc1}                          

\usepackage{graphicx}       

\usepackage{amsmath}        

\usepackage{cite}           

\volyear{XX}{2010}               
\journal{Can. J. Phys.}                       
\journalcode{cjp}                   
\filenumber{}                    

\received{}                      
\revreceived{}                   
\accepted{}                      
\revaccepted{}                   

\begin{document}





\title{Towards a first observation of magneto-electric directional anisotropy and linear birefringence in gases}

\author{C. Robilliard$^{(a,b)}$}
\address[label1]{$^a$ Universit\'e de Toulouse; UPS; Laboratoire collisions Agr\'egats R\'eactivit\'e, IRSAMC; F-31062 Toulouse, France. \\$^b$ CNRS; UMR 5589; F-31062 Toulouse, France.}

\correspond{email: cecile.robilliard@irsamc.ups-tlse.fr}

\author{G. Bailly$^{(a,b)}$}
\address[label1]

\shortauthor{C. Robilliard and G. Bailly} 

\maketitle

\begin{abstract}
   In this contribution to PSAS'2010 we report on recent progress on an experiment aimed at measuring small optical directional anisotropies by frequency metrology in a high finesse ring cavity. We focus on our first experimental goal, the measurement of magneto-electric effects in gases. After a review of the expected effects in our set-up, we present the apparatus and the measurement procedure, showing that we already have the necessary sensitivity to start novel experiments.
\end{abstract}

\begin{resume}
Dans cette contribution \`a PSAS'2010, nous pr\'esentons les avanc\'ees r\'ecentes sur une exp\'erience destin\'ee \`a mesurer de tr\`es faibles anisotropies optiques par m\'etrologie de fr\'equence dans une cavit\'e en anneau de haute finesse. Nous pr\'ecisons notre premier objectif exp\'erimental, la mesure d'effets magn\'eto-\'electriques dans les gaz. Apr\`es une revue des effets attendus dans notre montage, nous pr\'esentons l'appareil et la proc\'edure de mesure, montrant que nous avons d\'ej\`a la sensibilit\'e n\'ecessaire pour commencer des mesures novatrices.
\end{resume}


\section{Introduction}

Searching for anisotropies and, more generally, symmetry violations is one of the main trends of contemporary physics, with the aim of testing the present models and their proposed extensions. Let us cite as examples tests of Lorentz invariance \cite{Mattingly2005} or measurements of atomic parity violation \cite{Ginges2004}. For this purpose, low energy precision experiments appear to be complementary to high energy physics.

In this context where precise optical measurements are needed, high finesse optical cavities play an important role. Indeed, they allow for impressively sensitive apparatus, reaching relative sensitivities as low as $3\times 10^{-23} \; /\sqrt{\mathrm{Hz}}$ for km long gravitational waves interferometers \cite{LIGO,VIRGO} or $10^{-18}$ to $10^{-19} \; /\sqrt{\mathrm{Hz}}$ for table-top vacuum magnetic birefringence experiments \cite{Battesti2008,Berceau2010,DellaValle2010}.

In this paper we present the status of a new experiment aimed at measuring very small directional anisotropies thanks to a high finesse ring cavity. Our first objective is to measure effects induced in gases by magnetic and electric fields, which will allow us to optimize our experimental set-up. Subsequently, we should be able to measure these effects with enough precision to be sensitive to small contributions such as relativistic ones, and also to observe them in vacuum, which would be the first evidence of the nonlinearity of vacuum due to quantum effects. We start by describing these effects and the way we plan to measure them, after which we present our apparatus, a first version of it being now fully operational. We finally conclude by mentioning the near future measurements to be made and the improvements that are under way on the set-up.

\section{Optical anisotropies induced in gases by magnetic and electric fields}

In the 1980's, several authors performed some methodical studies of the optical anisotropies induced in material media, either solids or fluids, by electric and magnetic fields \cite{Baranova1977,Graham1983,Graham1984,Ross1989}. Special attention was devoted to effects bilinear in electric and magnetic fields, which had never been observed. Indeed, while the Kerr and Cotton-Mouton effects arise from lowest order interactions, namely electric dipole coupling, the bilinear magneto-electric effects originate from electric quadrupole and magnetic dipole couplings. Therefore they are considerably smaller, hence more difficult to observe. On the other hand, they provide molecular response functions that are not accessible to any other experiment.

Our project aims at measuring two of these effects in usual gases, starting with nitrogen N$_2$: magneto-electric directional anisotropy and magneto-electric linear birefringence. Both occur for crossed transverse magnetic fields.

\subsection{Directional anisotropy}

Magneto-electric directional anisotropy can be induced in all media, including centrosymmetric ones. It was first predicted by G.~E.~Stedman and coworkers \cite{Ross1989} and observed for the first time in a crystal by G.~L.~J.~A.~Rikken and coworkers \cite{Rikken2002}. This non-reciprocal effect is supposed to be independent on light polarization. Although it has never been calculated in material media, its order of magnitude is expected to be similar to that of the magneto-electric linear birefringence. 

\begin{figure}[htb]
\begin{center}
  \includegraphics[width=5cm]{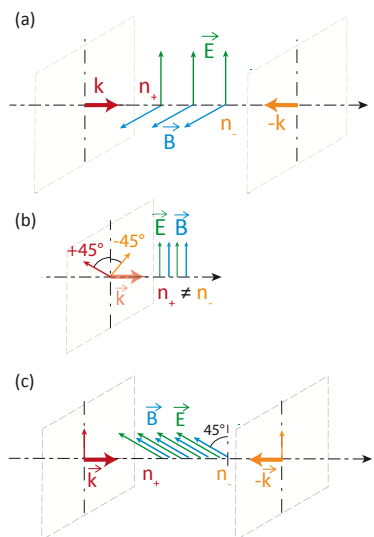}\\
  \caption{(color online) (a) Scheme of the magneto-electric directional anisotropy. The index of refraction of two waves counterpropagating perpendicularly to the crossed applied electric and magnetic fields are different. (b) Scheme of the magneto-electric Jones birefringence. (c) Scheme used in our experiment, converting the Jones birefringence into a directional anisotropy effect.}
  \label{fig:scheme}
\end{center}
\end{figure}

The scheme of this effect is presented in Fig. \ref{fig:scheme}(a). 

\subsection{Linear birefringence}

Similarly to directional anisotropy, the existence of linear birefringence is not submitted to any condition on the medium, and can therefore be induced in atomic or molecular gases by crossed transverse electric and magnetic fields. It has only been observed in molecular liquids \cite{Roth2002}, and never in gases. 

\begin{figure}[htb]
\begin{center}
  \includegraphics[width=6cm]{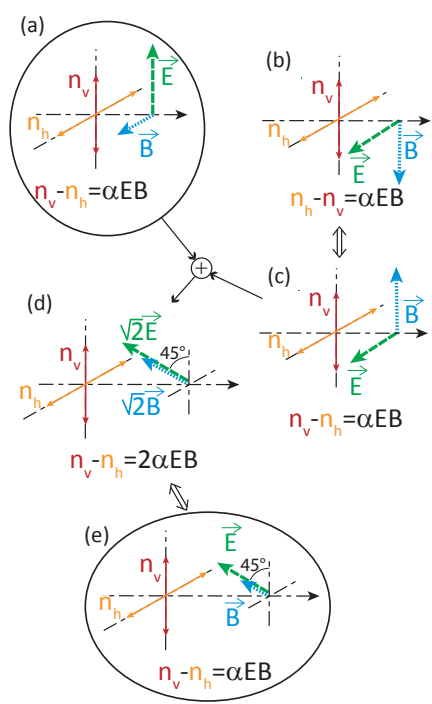}\\
  \caption{(color online) Graphical proof of the equivalence between magneto-electric linear birefringence (a) and magneto-electric Jones birefringence (e). (a) Scheme of linear birefringence, which is bilinear in $E$ and $B$. $n_v$ (resp. $n_h$) is the index of refraction of vertically (resp. horizontally) polarized light. (b) Situation after rotating $E$ and $B$ by $\pi/2$: $n_v$ and $n_h$ are simply inverted. (c) is deduced from (b) by $B$ inversion. (d) Superposition of situations (a) and (c). This is valid because of the bilinearity of the linear birefringence in $E$ and $B$, and also because there is no bilinear birefringence induced by parallel $E$ and $B$ fields with axis parallel and orthogonal to the fields \cite{Graham1984}. (e) After renormalising the fields, one obtains the Jones configuration, with parallel $E$ and $B$ and eigenaxis rotated by $\pm 45^\circ$ with respect to the fields.}
  \label{fig:Joneslinear}
\end{center}
\end{figure}

Although our apparatus is aimed at measuring directional anisotropies, we have found a scheme which is sensitive to linear birefringence. Actually, this scheme is based on the equivalence between magneto-electric linear birefringence in crossed fields and magneto-electric Jones birefringence, which occurs for parallel fields \cite{Graham1984}. Figure \ref{fig:Joneslinear} presents the symmetry arguments proving this equivalence. The proof is valid under the assumption that the system is centrosymmetric in the absence of external fields, and that the magneto-electric and Jones birefringences are bilinear in $E$ and $B$.

As a consequence, measuring the linear or the Jones birefringence is simply a matter of convenience. The scheme we are developing uses the Jones configuration and converts Jones birefringence into a directional anisotropy effect, as shown on Fig. \ref{fig:scheme}(b-c).

\subsection{Expected values}

There is not much literature on magneto-electric effects in gases. Jones birefringence has been computed in usual simple gases \cite{Rizzo2003} for non-resonant light, while analytical calculations are available for alcaline-earth atoms \cite{Mironova2006} and for alcaline atoms \cite{Chernushkin2008} in the vicinity of several atomic resonances. It is worth noting that for all the resonances studied, directional anisotropy is slightly larger than linear birefringence. Actually, both phenomena arise at the same development order of the interaction Hamiltonian and are therefore expected to have similar amplitudes. 

Both the magneto-electric directional anisotropy and the magneto-electric linear birefringence have been calculated in quantum vacuum \cite{Rikken2000,Rikken2003}, using the usual Euler-Heisenberg-Weisskopf effective Lagrangian.

We summarize in Table \ref{tab:calculation} the expected $\Delta n$ in fields $B=1\;$T and $E=1\;$V/m, for non-resonant light. For gases, the reference pressure is $P=1\;$bar and the temperature is $T=293\;$K. Conversion from values obtained in different temperature or pressure conditions is made within the ideal gas approximation. The calculations have been performed for $\lambda =632.8\;$nm, except for vacuum where it is valid for all wavelengths. We take into account the equivalence between linear and Jones birefringences: according to Fig. \ref{fig:Joneslinear}, we have $\Delta n_{\mathrm{Jones}}= n_{+45^\circ} -n_{-45^\circ}= n_B-n_E = \Delta n_{\mathrm{MELB}}$. The polarization angles for Jones birefringence are oriented counterclockwise from the fields direction as judged by an observer into whose eye the light is travelling; $n_B$ (resp. $n_E$) is the refraction index in the $B$ (resp. $E$) direction. Concerning directional anisotropy, we define $\Delta n_{\mathrm{MEDA}}=n_{+} - n_{-}$ where the positive propagation direction is along $\mathbf{E}\times \mathbf{B}$. 

\begin{table}[htb]
\begin{center}
  \begin{tabular}{|l|l|c|c|}
  \hline {\bf Medium} & {\bf Ref.} & $\eta_{\mathrm{MELB}}$ (T$^{-1}$.V$^{-1}$.m) & $\eta_{\mathrm{MEDA}}$ (T$^{-1}$.V$^{-1}$.m) \\ \hline \hline
  Quantum Vacuum & \cite{Rikken2000,Rikken2003} & $2.7\times 10^{-32}$ & $-6.7\times 10^{-32}$ \\ \hline
  Helium He & \cite{Rizzo2003} & $1.6\times 10^{-24}$ & - \\ \hline
  Hydrogen H & \cite{Graham1983} & $3.4\times 10^{-23}$ & - \\ \hline
  Neon Ne & \cite{Rizzo2003} & $4.2\times 10^{-24}$ & - \\ \hline
  Argon Ar & \cite{Rizzo2003} & $3.6\times 10^{-23}$ & - \\ \hline
  Krypton Kr & \cite{Rizzo2003} & $7.8\times 10^{-23}$ & - \\ \hline
  Hydrogen H$_2$ & \cite{Rizzo2003} & $4.8\times 10^{-23}$ & - \\ \hline
  Nitrogen N$_2$ & \cite{Rizzo2003} & $9.0\times 10^{-23}$ & - \\ \hline
  Carbon monoxide CO & \cite{Rizzo2003} & $1.4\times 10^{-22}$ & - \\ \hline
  \end{tabular}
  \caption{Calculated magneto-electric linear birefringence $\eta_{\mathrm{MELB}} = n_B-n_E$ and magneto-electric directional anisotropy $\eta_{\mathrm{MEDA}}= n_{+} -n_{-}$ in gases and in vacuum, normalized to $B=1\;$T, $E=1\;$V/m. For gases, the reference pressure and temperature are $P=1\;$bar and $T=293\;$K. The calculations were made at $\lambda =632.8\;$nm.}\label{tab:calculation}
\end{center}
\end{table}

The numbers in Table \ref{tab:calculation} clearly show that measurements in gases are not beyond present technological possibilities, but are nevertheless challenging. Indeed, using a magnetic field on the order of 10~T and an electric field on the order of 10~MV/m one gets $\Delta n\simeq 10^{-14}\;\mathrm{to}\; 10^{-16}$ in gases and $\Delta n\simeq 7\times 10^{-24}$ in quantum vacuum!

Our first measurement will be the magneto-electric directional anisotropy of Nitrogen, a gas which should exhibit a large effect while being easy to handle. 

In the following, we describe our apparatus.

\section{Experimental set-up}

\begin{figure}[htb]
\begin{center}
  \includegraphics[width=8cm]{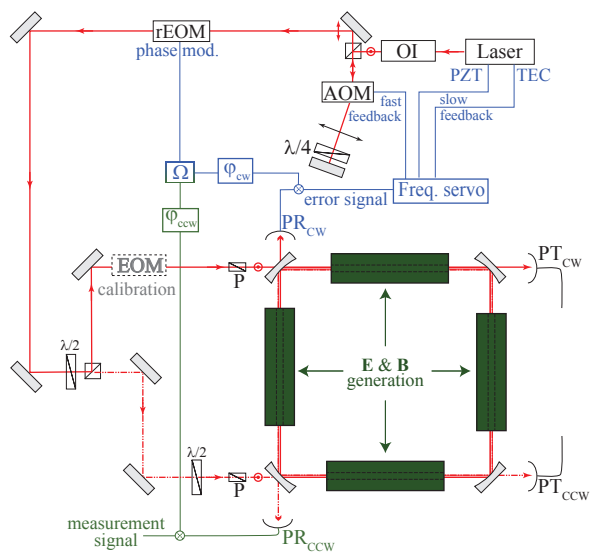}\\
  \caption{Experimental set-up of our experiment (color online). The beampath is represented in red with arrows indicating the propagation direction, the frequency stabilization system in blue and the measurement signal generation in green. The green blocks on each cavity arm represent the rods generating the magnetic and electric fields for magneto-electro-optic measurements in gases. The grey EOM inserted in the clockwise beam path just before the cavity allows to calibrate the experiment. An optical isolator (OI) prevents feedback noise; the laser beam frequency is then frequency-shifted with an acousto-optic modulator (AOM) in a cat's eye retroreflector. A resonant electro-optic modulator (rEOM) provides the phase modulation at frequency $\Omega$ for the Pound-Drever-Hall frequency stabilization. The servo actuators are the laser thermo-electric cooler (TEC), the laser piezo-electric transducer (PZT) and the AOM. The light polarization is controlled all along the beampath by halfwave ($\lambda/2$) and quarterwave ($\lambda/4$) retardation plates, and by polarizers (P). Light is injected into the cavity both in the clockwise (cw) and counterclockwise (ccw) directions; the $\mathrm{PR_{cw}}$ and $\mathrm{PR_{ccw}}$ (resp. $\mathrm{PT_{cw}}$ and $\mathrm{PT_{ccw}}$) photodiodes monitor the reflected (resp. transmitted) power in both directions.}
  \label{fig:setup}
\end{center}
\end{figure}

Our apparatus is described in details in \cite{Bailly2010}. The experiment, represented in Fig. \ref{fig:setup}, mainly consists of a square resonant cavity of total length $L=4L_0=1.6\;$m, with a finesse varying between $15000$ and $50000$ depending on mirror pollution. 

\subsection{Optical set-up}

Two light beams of identical frequency, originating from the same Nd:YAG laser, are injected into the cavity, respectively clockwise and counterclockwise. Their frequency is stabilized on the cavity resonance for the clockwise laser beam, following the well-known Pound-Drever-Hall scheme \cite{Drever1983,Black2000}. An error signal similar to the clockwise one is generated in the counterclockwise direction: provided the frequency stabilization is good enough, this error signal is proportional to the resonance frequency difference between the two counterpropagating beams. In order to improve the signal-to-noise ratio, we implement a homodyne detection scheme by modulating the effect to be measured at frequency $f_{\mathrm{mod}}$ and using a lock-in amplifier.

Additionnally, both beams are superimposed nearly all along and thus experience the same mirror vibrations and the same fluctuations of the propagating medium. We have checked that the spectra of both error signals are identical, and the noise power spectral density in the relevant frequency range (a few hundreds Hertz) is less than an order of magnitude over shot-noise. Very low noise frequency stabilization and efficient common mode noise rejection are the two key elements explaining the good sensitivity of our apparatus even though it is not yet in a vacuum chamber.

As explained in \cite{Bailly2010}, we have performed sensisitivity tests thanks to a broadband electro-optic modulator (EOM) placed on the clockwise beam just before the cavity and supplied with a sinusoidal voltage at $f_{\mathrm{mod}}$. We have checked that the amplitude of the error signal frequency component at $f_{\mathrm{mod}}$ is proportional to the amplitude of frequency modulation, and that the apparatus is able to detect frequency modulations as low as $200\;\mu$Hz rms.

\subsection{External fields generation}

We have built 4 rods generating crossed transverse magnetic and electric fields. These rods are $L_{E\times B}=20\;$cm long and have a $4\times 4\;$mm$^2$ hole in their center for optical access.

\begin{figure}[htb]
\begin{center}
  \includegraphics[width=8cm]{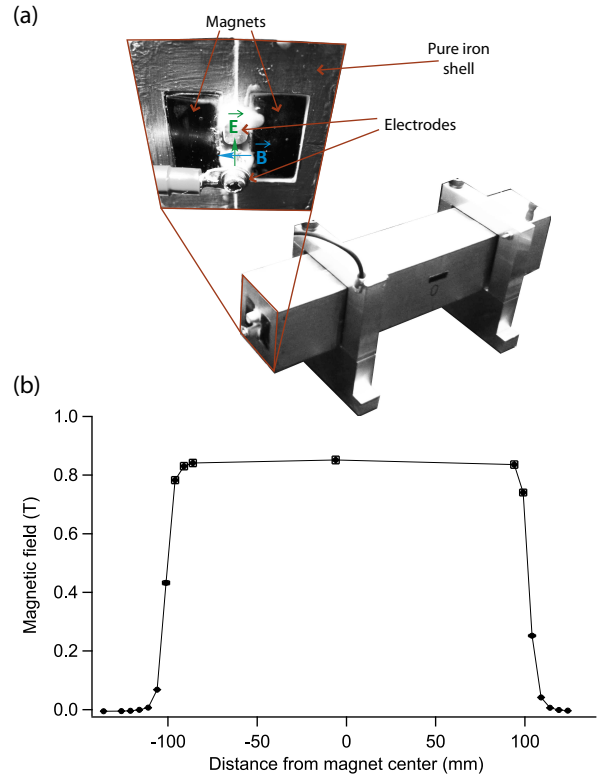}\\
  \caption{(color online) (a) Arrangement of the rods generating the electric and magnetic fields. The hole for optical access is $4\times 4\;$mm$^2$. (b) Profile of the magnetic field along the optical path. The field well inside the magnet is $0.185\;$T, as well as the effective field defined as $B_{\mathrm{eff}} = \int{B\,\mathrm{d}l}/L_{E\times B}$.}
  \label{fig:EBrods}
\end{center}
\end{figure}

The magnetic field is generated by two permanent NdFeB block magnets with dimensions $200\times 20\times 10\;$mm$^3$,  surrounded by a pure iron shell in order to guide the field lines and to enhance $B$ between the two opposite magnets, as illustrated in Fig. \ref{fig:EBrods}(a). We have measured the magnetic field profile along the beam path with a Hall effect probe: it is presented if Fig. \ref{fig:EBrods}(b). All the 4 rods provide the same magnetic field within less than $2\%$. The magnetic field is uniform and equal to $0.185\;$T within $2\%$ in the central $185\;$mm, and rises from $1\%$ to $98\%$ of its maximum value in $20\;$mm. In the transverse direction, moving 1 mm away from the center of the hole induces a field increase of $1\%$. To avoid this we adjust the rods centering after each cavity realignment. 

The electric field originates from a pair of aluminum electrodes with a roughly $4\times 4\;$mm$^2$ square section, the angles being rounded to prevent breakdowns. They are straight and parallel within $0.3\;$mm. These electrodes are supplied with a sine wave voltage through a $\pm 2\;$kV high voltage amplifier. They are $4\;$mm apart, symmetrically arranged with respect to the light path. The electrodes are glued between the magnets with an isolating epoxy resin. During operation, one of the electrodes is grounded while the other is connected to high voltage. The electrodes can be switched easily to reverse the electric field direction. We operate the electrodes between 200 and 500 Hz, and we have checked that the voltage distortion is always negligible at these frequencies, even when the four rods are supplied simulatneously at the maximum voltage.

Finally, our rods can provide fields $B_{\mathrm{eff}}=0.85\;$T and $E=0.35\;$MV/m rms on a total length $4L_{E\times B}=\frac L2 =0.8\;$m. In such fields, the order of magnitude of the expected frequency modulation due to Jones birefringence in Nitrogen (see Table \ref{tab:calculation}) is $\delta n=2.7\times 10^{-17}\;$rms, hence 

\begin{eqnarray}
\delta \nu & = & \nu \frac{4L_{E\times B}}{L} \delta n \\
 & = & 2.8\times 10^{14} \times \frac 12 \times 2.7\times 10^{-17} \nonumber \\
\delta \nu & = & 3.8\;\mbox{mHz rms}
\end{eqnarray}

Since our rods provide crossed fields rather than parallel ones as required for Jones birefringence, we will measure the magneto-electric directional anisotropy in Nitrogen, and we therefore expect a slightly different value, probably a few times higher as far as we can infer from various calculations \cite{Rikken2000,Rikken2003,Mironova2006,Chernushkin2008}. Comparison with the present $200\;\mu$Hz sensitivity of our apparatus makes us confident that we should achieve soon the first observation of magneto-electric directional anisotropy in a gas.

\subsection{Measurement procedure}

The procedure we follow for the measurements uses the broadband EOM to calibrate the error signal before and after each run. Indeed, the conversion factor between the amplitude of the error signal frequency component at $f_{\mathrm{mod}}$ and the frequency modulation at this same frequency is proportional to the intracavity power, which varies according to cavity misalignment and mirror pollution. On our experiment, it decreased by an order of magnitude on a time scale of three months without any cavity realignment nor mirror cleaning.

\begin{figure}[htb]
\begin{center}
  \includegraphics[width=6cm]{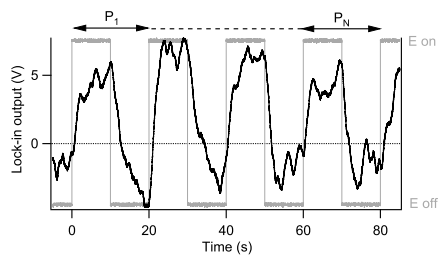}\\
  \caption{Typical signal, obtained for a frequency modulation of $6.5\;$mHz rms applied to the calibration EOM. We periodically switch on and off the electric field, with a period $T_{\mathrm{AM}}$. The signal processing consists of averaging the difference between the signal with the electric field on and that with the field switched off over the $N$ periods.}
  \label{fig:signal}
\end{center}
\end{figure}

Besides, we periodically switch on and off the electric field, with a period $T_{\mathrm{AM}}$ of a few tens of seconds, to avoid long term drifts of the zero field signal. Presently, our signal processing simply consists of calculating for each period $P_i$ (see Fig. \ref{fig:signal}) the difference $\delta V_i$ between the signal with $E$ on and the one with $E$ off, excluding the settling time due to the lock-in amplifier time constant after each electric field switch. Then we simply average over the $N$ periods: $\delta V=\frac 1N \sum_{i=1}^N \delta V_i$. 

This way, we have reached a relative frequency sensitivity about $10^{-16}/\sqrt{\mbox{Hz}}$. In terms of the expected Jones birefringence $\delta n$ in Nitrogen, this amounts to around $10\,\delta n/\sqrt{\mbox{Hz}}$. 

\section{Outlook and conclusion}

In conclusion, we have shown that our apparatus is now fully operational and should provide us in the near future with a measurement of magneto-electric directional anisotropy in gases. At present, we can make only measurements in harmless gases such as air or nitrogen, since our optical cavity is not yet in a tight gas chamber.

The next steps in our experimental effort will be on the one hand the installation of the high finesse cavity in a vacuum chamber. This will not only allow us to perform measurements in various gases, but it will also avoid mirror pollution, which is presently the main reason for excess noise over the quantum limit. This will also bring some improvements in thermal and acoustic insulation. Along with some work concerning the cavity finesse, the optical coupling and the frequency stabilization, this step will allow us to reach a near shot-noise limited relative frequency sensitivity below $10^{-20}/\sqrt{\mbox{Hz}}$. On the other hand, we are designing new rods with higher magnetic and electric fields, generating either crossed or parallel fields close to 2 T and to 3 MV/m. Such an apparatus will open the way to very precise measurements of directional anisotropies.

On the longer term, we want to use even higher fields, in order to observe the magneto-electric directional anisotropy in quantum vacuum, which is an experimental challenge. With fields  $B=15\;$T and $E=20\;$MV/m, already demonstrated in other groups, the signal-to-noise ratio would reach unity within 10 h.

Besides, our measurements in the absence of external fields may also provide some tests of Lorentz invariance and constrain some of the proposed extensions to Standard Model.

\end{document}